\documentclass[%
preprint,
 amsmath,amssymb,
 prl,
]{revtex4-1}

\usepackage{graphicx}
\usepackage{dcolumn}
\usepackage{bm}
\usepackage{mathrsfs}
\usepackage{rotating}
\usepackage{threeparttable}
\def\comment#1{}
\usepackage{color}

\begin{document}

\preprint{APS/123-QED}

\title{Accurately recalibrated waveforms for extreme-mass-ratio inspirals in effective-one-body frame}

\author{Ran Cheng $^{1,2}$ }
 \author{  Wen-Biao Han $^{1,2}$}
 \email{wbhan@shao.ac.cn}
\affiliation{%
 1. Shanghai Astronomical Observatory, Shanghai, 200030, China   \\
 2. School of Astronomy and Space Science, University of Chinese Academy of Sciences,  Beijing 100049, China \\
}%





\begin{abstract}
How to calculate the gravitational waves (GWs) of Extreme-mass-ratio-inspirals (EMRIs) in a highly accurate and efficient way still keeps a challenge. In this paper, we present a so-called fully recalibrated waveforms for EMRIs with high accuracy. Based on the numerical data by solving the Teukolsky equations, we recalibrate all mass-ratio independent coefficients of the factorized waveforms which are used in the effective-one-body (EOB) models. Due to these new coefficients, the precision of waveforms is improved enormously, and is much higher than the original forms and at the same time  higher than other existing calibration models. We believe our model will play an important role in the waveform-template construction of the space-based GW detectors.
\end{abstract}

\pacs{Valid PACS appear here}
\maketitle
\section{Introduction}
Extreme-mass-ratio-inspirals (EMRIs) are composed by supermassive black holes and stellar-mass compact bodies. For example, an EMRI event is that a compact body (stellar massive black hole, neutron star etc.) orbits around a supermassive black hole in a strong field zone, and the radius of orbit shrinks due to the radiation of gravitational waves (GWs).  People popularly believe that there are supermassive blackholes reside in the centre of most galaxies \cite{alexander05}. The typical mass of a supermassive blackhole ranges from about $10^5$ to $10^7$ solar masses ($M_\odot$). The frequency of GWs from EMRIs then can be around $10^{-2} \sim 10^{-1}$ Hz when the small body's orbit is in the wildly relativistic region of the central black hole. EMRIs are very important gravitational wave sources for the space-based detectors like as eLISA\cite{elisa}, Taiji\cite{taiji} and Tianqin\cite{tianqin}. From the estimations of \cite{finn00, cutler02,amro07,barack03,gair04}, eLISA should see about 50 such EMRI events per year out to redshifts $z \approx 1$, based on calculations of stellar dynamics in galaxies' central cusps\cite{hopman06}.

As the same as LIGO, these space-based detectors also adopt the matched-filtering technology to search the GW signals, and analyze the properties and parameters of sources. This requires a huge number of waveform templates in varied parameter space. How to calculate the waveforms of EMRIs  in a highly accurate but efficient way still keeps a challenge. Usually, one can calculate the waveforms accurately by numerical simulations (for example, numerically computing the blackhole perturbation equations \cite{rwz1,rwz2,teukolsky1,teukolsky2}), but the computing speed is too slow to generate enough templates. Or alternatively, one can use post-Newtonian approximation (including the hybrid method, see e.g. \cite{gair06,babak08,carlos11}) to generate the waveforms with a fast speed but will lost the accuracy in the strong field.

However, considering the long time-scale of the waveforms of EMRIs, the requirement for accuracy and efficiency of waveforms is quite high. In the final year of the inspiral, an EMRI waveform has $10^5$ circles \cite{gair04}. If we limit the dephase to be less than one circle between the theoretical templates and the real waveforms, the accuracy of the radiation reaction should be at $10^{-5}$ level. At the same time, the inspiral waveform depends on 14 different parameters \cite{glampedakis02}. Based on the analysis of \cite{gair04}, if we assume that only about 8 of these 14 parameters affect the phase evolution, it will need $ 10^{5\times 8}$ templates to perform a fully coherent matched-filtering search for year-long inspiral waveforms.

In the LIGO's GW searching, they use effective-one-body (EOB) model \cite{eob14,eob16} to construct waveform templates. \cite{ligo16}. These EOB waveforms are calibrated by numerical relativity data to guarantee the accuracy. Because the EOB method is basically an analytical post-Newtonian (PN) formalism, the computing efficiency is very high. The EOBNR model works well in the comparable mass-ratio binaries, but may not in the EMRIs. Nevertheless, this inspires us to develop a similar waveform model in the EOB frame, i.e., calibrating the waveform formalisms by the numerical Teukolsky-based data. However, the first attempt has been done by Yunes et. al. \cite{Yunes10,Yunes11}. They calibrated out several higher PN order coefficients to improve the accuracy of the factorized-resummed waveforms. In the present work, we recalibrate all the coefficients in the factorized-resummed waveform formalisms and add a few higher order coefficients.  In this sense, we call our method as a fully recalibration model. We will see the accuracy of the waveforms improved much more by our new recalibration .

The paper is organized as follows. In the next section, we introduce the original PN waveforms in the EOB frame. A brief summation of the Teukolsky equations and numerical algorithm is given in the section III. We then discuss our full recalibration method, results and comparisons in the section IV. At last, conclusions and remarks are given. Throughout the paper, we use units $G=c=1$ and the metric signature $(-,+,+,+)$. Distance
and time are measured by the central black-hole mass $M$.

\section{The factorized-resummed EOB waveforms}
To overcome the very poor convergence of the PN series (especially near the ISCO, i.e. innermost stable circular orbit) pointed out by Cutler et al. \cite{cutler93} and Poisson \cite{poisson95}, Damour et. al. suggested to use resummation methods to extend the numerical validity of the PN expansions (at least) up to the ISCO \cite{damour98}.  This technology was improved constantly (for example, see \cite{Damour07,Damour09,panyi11}), and it formed the factorized-resummed PN waveforms which aims at extending the validity of suitably resummed PN results beyond the ISCO, and up to the merger. The waveforms were later introduced in the EOB approach and used to construct the complete GW templates.

The factorized-resummed modes read
\begin{align}\label{hlm}
h_{lm}=h_{lm}^{\text{Newt,}\epsilon_p}S^{\epsilon_p}_{lm}T_{lm}e^{i\delta_{lm}}(\rho_{lm})^l\,,
\end{align}
where $\epsilon_p$ indicates the parity of the multipolar waveform, and $h^{Newt,\epsilon_p}_{lm}$ is the Newtonian part, $S^{\epsilon_p}_{lm}(v)$,$T_{lm}(v)$,$\delta_{lm}(v)$ and $\rho_{lm}$, the terms appearing in Eq. (\ref{hlm}), we could refer to Reference\cite{panyi11,Tara12} for detailed expressions. Especially, $\rho_{lm}$ is a PN expansion about $(v/c)^n$, the numeric coefficients can be divided into four groups: mass-ratio and spin independent, mass-ratio dependent, spin dependent, and mass-ratio and spin both dependent coefficients.

Then the energy fluxes down to the infinity $\dot E^{\infty}$ can be expressed with the harmonically-decomposed waveforms $h_{lm}$ by
\begin{align}
\dot E^{\infty}=\sum\limits_{l=2}^{8} \sum_{m=1}^{l} \frac {\omega^2_m\left|{h_{lm}}\right|^2}{8\pi}, \label{flux8}
\end{align}
in this expression, $\omega_m=m\Omega$, where $\Omega$ is the orbital frequency. Now the PN waveforms are only expanded to multipoles $l=8$.

In the EMRI cases, one can also add the energy fluxes down to event horizons \cite{Yunes10}. We calculate the flux through the N/2-th order Taylor expansion in PN theory\cite{PJA05}
\begin{align}\label{eh}
 \dot{E}^{\rm H}=32/5\nu^2 v^{10} \sum\limits_{n=0}^N \left[{a_n(\nu)+b_n(\nu)\log(v)}\right]v^n,
\end{align}
where $\nu$ is the symmetric mass-ratio which equals to $m_1m_2/(m_1+m_2)^2$ ($m_1$ and $m_2$ are the masses of the two components), and $v$ is the circular orbit frequency. Following reference\cite{Yunes10}, we classify the PN parameters based on their physical origin and condition of spin inclusion. Those that correspond to radiation in falling into the horizon will be labeled ($a_n^{\text{Hor}},b_n^{\text{Hor}}$), with a superscript S (NS) if they are further spin-dependent(spin-independent), where $a^{\text{Hor,NS}}_{<14}$, $b^{\text{Hor,NS}}_n$, $a^{\text{Hor,S}}_{<11}$ and $b^{\text{Hor,S}}_{<11}$ have been specified in reference\cite{Yunes10}. For more details, see reference\cite{Mino97}. 

\section{The fully recalibrated waveforms}
Until now, the analytical forms of $\rho_{lm}$ are expanded up to 5PN order, for example
\begin{align}
\rho_{22} = 1+  (\frac{55}{84} \nu - \frac{43}{42}) v^2 -\frac{2}{3} q(1-\frac{\nu}{2}) v^3 + \cdots \,,
\end{align}
where $q$ is the dimensionless Kerr parameter $q\equiv J/M^2$, and $J$ is the angular momentum of the black hole. As already shown in Ref. \cite{Yunes10}, the accuracy of these original waveforms is far away from the requirement in the EMRI simulations. Yunes et.al. calibrated a few of higher order coefficients to improve the accuracy of the resummed formalisms.  In the previous work \cite{han16} by one of us, we fitted out a set of polynomials from the Teukolsky-based fluxes to calculate $\dot{E}^{\infty, \text{H}}$ and waveforms accurately. In the present work, we  keep the forms of the factorized-resummed expansions, but recalibrate all numeric coefficients of these expansions. The reference data for the recalibration are calculated by the highly accurate solutions of the Teukolsky equations \cite{teukolsky1,teukolsky2}. The details of the Teukolsky equations and our numerical algorithm are addressed in our previous papers \cite{han10,han11,han14} and the references inside.

For the recalibration, following Ref. \cite{Yunes10,Yunes11,panyi11,Damour09}, we first write down the expressions of the $\rho_{lm}$ with undetermined numeric coefficients,
for $m = l$,
\begin{align} \label{le2m}
\rho_{lm}=&1+a_1 v^2+b_1q v^3
+(a_2+b_2q^2)v^4+b_3qv^5+[b_4q^2+a_3+a_4{\rm eulerlog}(mv)]v^6 \nonumber\\
&[b_{5}q+ b_{6}q^3]v^7+[b_7q^2+b_8q^4+a_5+a_6{\rm eulerlog}(mv)]v^8+ (b_9q+b_{10}q^3)v^9 +\nonumber\\
&[a_7+a_8{\rm eulerlog}(mv)]v^{10}+[a_9+a_{10}{\rm eulerlog}(mv)]v^{12},
\end{align}
where $\text{eulerlog}(mv)=\gamma_E+\log(2mv)$, $\gamma_E$ = 0.577215... is Euler’s constant, and for $m \neq l$,
\begin{align} \label{lne2m}
\rho_{lm}=&1+b_1qv+({a_1+b_2q^2})v^2+(b_3q+b_4q^3)v^3
+(a_2+b_5q^2+b_6q^4)v^4+ \nonumber\\
&(b_7q+b_8q^3+b_9q^5)v^5+[a_3+a_4{\rm eulerlog}(mv)+b_{10}q^2+b_{11}q^4+b_{12}q^6]v^6+ \nonumber\\
&[b_{13}q+b_{14}{\rm eulerlog}(mv)q+b_{15}q^3+b_{16}q^5+b_{17}q^7]v^7+[a_5+a_6{\rm eulerlog}(mv)]v^8+ \nonumber\\
&[a_7+a_8{\rm eulerlog}(mv)]v^{10}+[a_9+a_{10}{\rm eulerlog}(mv)]v^{12}.
\end{align}
The coefficients $a_1, ~a_2, ~a_3, \cdots$ are spin-independent, and $b_1, ~b_2, ~b_3, \cdots$are spin-dependent. Considering the small mass-ratio limit, all terms that depend on mass-ratio have been omitted.

To get all the numeric coefficients in Eqs. (\ref{le2m},~\ref{lne2m}), we adopt a two-dimensional (i.e., in spin parameter $q$ and orbital velocity $v$) least-squares method to fit them. The data for the least-squares fitting come from the highly accurate Teuksolksy-based fluxes.  We have 17 group of numerical data with varied spin parameters $q$: -0.9, -0.7, -0.5, -0.3 -0.1, 0.0, 0.1, 0.2, 0.3, 0.4, 0.5, 0.6, 0.7, 0.75, 0.8, 0.85, 0.9. Where positive q means that the direction of spin of the Kerr black hole is in parallel with orbital angular momentum of the small object, and negative q means anti-parallel case.  Because the ISCO is closer to the black hole when $q$ approaches 1, and then the variations of fluxes are wilder, we insert two points $q= 0.75,~0.85 $ between 0.7 and 0.9. For each $q$, we sample 18 points in the orbital velocity parameters, from $v \approx 0.02 \sim 0.08$ (depends on $q$) to the velocity at ISCO $v_\text{isco}$. These 18 points are generated by the Chebyshev nodes to obtain better performance on the boundary.

Therefore we totally have $17 \times 18 = 306$ sampling points on the two dimensional parameter space $\{q, ~v\}$.  Combining Eqs. (\ref{hlm}) and Eq. (\ref{flux8}), for each $l m$ mode, one has
\begin{align}
\rho_{lm} = \left[\frac{\dot{E}^\infty_{lm}}{\omega^2_{m} |h^{{\rm Newt}, \epsilon_p} S^{\epsilon_p}_{lm}T_{lm}|}\right]^{1/l} \,.
\end{align}
The precise values of fluxes $\dot{E}_{lm}^{\infty}$ on these sampling points are provided by the numerical calculation of the Teukolsky equations. $\omega_{m} = m \Omega$, and $\Omega$ can be calculated from the relation $ v = (M\Omega)^{1/3}$. All the other terms in the denominator can be directly calculated from the analytical expressions. The value of right hand now is known, then one can fit out all coefficients in Eqs. (\ref{le2m}, \ref{lne2m}) by using the two-dimensional least-squares method. All coefficients in $\rho_{lm}$ are listed in Tables \ref{gcoeflm}-\ref{coeflnm3}.

\begin{sidewaystable}
\centering
\begin{threeparttable}[b]
\caption{The coefficients of $m = l$ modes which are valid for arbitrary spin parameters.}
\label{gcoeflm}
\footnotesize
\begin{tabular}{c| c c c c c c c }
\hline \hline
\hspace{-1in}$ $&$l=2,m=2$&$l=3,m=3$&$l=4,m=4$&$l=5,m=5$&$l=6,m=6$&$l=7,m=7$&$l=8,m=8$
 \\

\hline
$a_1$   &~-1.022468181715E+00	&~ -1.166063960195E+00 	&~ -1.2223563676E+00	&~ -1.2484467258E+00	&~ -1.2616929716E+00	&~ -1.2687342542E+00	&~-1.2725151843E+00 \\
$a_2$   &~-2.556630282274E+00	&~ -1.982233742477E+00 	&~ -1.7977680536E+00	&~ -1.7166118773E+00	&~ -1.6711457762E+00	&~ -1.6429137727E+00	&~-1.6236873048E+00 \\
$a_3$   &~-2.989362917774E+01	&~ 1.777643876062E+01	 &~ 2.8862471815E+01	&~ 3.3375643363E+01	&~ 3.5773836101E+01	&~ 3.7349039618E+01	&~3.8475937921E+01  \\
$a_4$   &~-1.321863979254E+02	&~ -6.496732141750E+01 	&~ -4.3593643501E+01	&~ -3.4584344713E+01	&~ -2.9436736647E+01	&~ -2.6179251607E+01	&~-2.3903230730E+01 \\
$a_5$	  &~ 3.003861000861E+03	&~ 2.217282903432E+03	 &~ 1.8065494119E+03	&~ 1.6222940805E+03	&~ 1.5079148776E+03	&~ 1.4343768474E+03	&~1.3822088602E+03  \\
$a_6$	  &~-3.890299900936E+03	&~ -1.898295906580E+03 	&~ -1.2426002760E+03	&~ -9.6933498883E+02	&~ -8.1398529924E+02	&~ -7.1594899499E+02	&~-6.4772285890E+02 \\
$a_7$	  &~ 2.834727515529E+04	&~ 1.726656884980E+04	 &~ 1.2828836431E+04	&~ 1.0959187135E+04	&~ 9.8657732260E+03	&~ 9.1763133948E+03	&~8.6976470824E+03  \\
$a_8$	  &~-1.639831498106E+04	&~ -8.112548558898E+03 	&~ -5.3071933334E+03	&~ -4.1509912011E+03	&~ -3.4958283397E+03	&~ -3.0834244852E+03	&~-2.7973654386E+03 \\
$a_9$	  &~ 2.440335911505E+04	&~ 1.381920273654E+04	 &~ 9.7823787429E+03	&~ 8.1150627832E+03	&~ 7.1544600843E+03	&~ 6.5497861178E+03	&~6.1309435990E+03  \\
$a_{10}$&~-7.977993019971E+03	&~ -3.997260949304E+03 	&~ -2.6118990642E+03	&~ -2.0450559353E+03	&~ -1.7238444361E+03	&~ -1.5216640744E+03	&~-1.3815206406E+03 \\
$b_{1 }$&~-6.654444310941E-01	&~ -6.657707852251E-01 	&~ -6.6550546520E-01	&~ -6.6535003174E-01	&~ -6.6521707979E-01	&~ -6.6510326627E-01	&~-6.6500448915E-01 \\
$b_{2 }$&~ 4.839556166281E-01	&~ 4.921897888084E-01	 &~ 4.9572739626E-01	&~ 4.9775999773E-01	&~ 4.9906391726E-01	&~ 5.0000774263E-01	&~5.0074664313E-01  \\
$b_{3 }$&~-1.596446922126E+00	&~ -1.309300627500E+00 	&~ -1.2500372399E+00	&~ -1.2465280316E+00	&~ -1.2608775182E+00	&~ -1.2804955637E+00	&~-1.3008795759E+00 \\
$b_{4 }$&~ 5.734839505377E-01	&~ 2.283136207085E-01	 &~ 9.3277944248E-02	&~ 2.1277951567E-02	&~ -2.0841789689E-02	&~ -4.8607192848E-02	&~-6.8572937246E-02 \\
$b_{5 }$&~ 1.789698922772E+00	&~ 2.019415567476E+00	 &~ 2.2718827690E+00	&~ 2.4328006497E+00	&~ 2.5456622830E+00	&~ 2.6312303836E+00	&~2.6993620269E+00  \\
$b_{6 }$&~-4.327512994411E-01	&~ -4.697274956101E-03 	&~ 1.0374681155E-01	&~ 1.4977614531E-01	&~ 1.7279574323E-01	&~ 1.8550158778E-01	&~1.9301428447E-01  \\
$b_{7 }$&~ 1.439338856421E+00	&~ -2.034145921637E-01 	&~ -3.3067312867E-01	&~ -2.5175639083E-01	&~ -1.4687292565E-01	&~ -3.8918784264E-02	&~6.4618405437E-02  \\
$b_{8 }$&~-8.219009282843E-01	&~ -3.369305813079E-01 	&~ -2.5338866309E-01	&~ -2.1292821492E-01	&~ -1.8954553725E-01	&~ -1.7498564232E-01	&~-1.6512204150E-01 \\
$b_{9 }$&~-1.945096581021E+01	&~ -1.153450623209E+01 	&~ -1.0487653390E+01	&~ -1.0433322943E+01	&~ -1.0636532049E+01	&~ -1.0926117902E+01	&~-1.1238490218E+01 \\
$b_{10}$&~ 7.257863585947E+00	&~ 3.789085707619E+00	 &~ 2.9438753159E+00	&~ 2.5549894720E+00	&~ 2.3444258931E+00	&~ 2.2164508614E+00	&~2.1316479097E+00  \\

\hline \hline
\end{tabular}

  \begin{tablenotes}
    \item [a] { }
   \end{tablenotes}
\end{threeparttable}
\end{sidewaystable}
\clearpage

\begin{table}[h!]
\caption{The coefficients of $m \neq l$ modes which are valid for arbitrary spin parameters.}
\label{coeflnm1}
\begin{tabular}{c| c c c c  }
\hline \hline

$ $&$l=2,m=1$&$l=3,m=2$&$l=3,m=1$&$l=4,m=3$
 \\
\hline
$a_1 $  &~~ -1.0527707685E+00 &~~ -1.2145240791E+00   &~~ -7.21524363E-01    &~~ -1.2612372247E+00   \\
$a_2 $  &~~ -9.3241462756E-01 &~~ -8.5923599933E-01   &~~ -3.11113895E-01    &~~ -8.9933197223E-01   \\
$a_3 $  &~~ -1.6759078476E+01 &~~ 3.1951845221E+00    &~~ -7.48783760E+01    &~~ 4.5707364494E+00    \\
$a_4 $  &~~ -1.8669253387E+01 &~~ 2.1653247307E+00    &~~ -7.29364137E+01    &~~ 1.2609902334E+01    \\
$a_5 $  &~~ -5.6495769926E+01 &~~ -1.8136285648E+01   &~~ 1.07123233E+02     &~~ -3.9533270552E+02   \\
$a_6 $  &~~ -7.0266261584E+02 &~~ -1.6087231421E+02   &~~ -2.35621409E+03    &~~ 2.4225430221E+02    \\
$a_7 $  &~~ 3.5829757446E+03  &~~ 2.6077841852E+03    &~~ 1.06529090E+04     &~~ -7.2699350331E+02   \\
$a_8 $  &~~ -4.0723679649E+03 &~~ -1.8307910870E+03   &~~ -1.06104376E+04    &~~ 1.5064300121E+02    \\
$a_9 $  &~~ 5.8959149522E+03  &~~ 4.0759893996E+03    &~~ 1.27864527E+04     &~~ 8.6135380084E+02    \\
$a_{10}$&~~ -2.6369522589E+03 &~~ -1.4161892952E+03   &~~ -5.46673244E+03    &~~ -2.9976594051E+02   \\
$b_{1	}$&~~ -7.4984593296E-01 &~~ -4.4442747465E-01   &~~ 1.48939169E-05     &~~ -3.1252866834E-01   \\
$b_{2	}$&~~ -2.8398487091E-01 &~~ -1.9938456069E-01   &~~ -1.48791360E-04    &~~ -1.4770551822E-01   \\
$b_{3	}$&~~ 1.7471030805E+00  &~~ 9.5894558412E-01    &~~ -1.85529718E+00    &~~ 5.6157612198E-01    \\
$b_{4	}$&~~ -2.0583114811E-01 &~~ -1.4524057397E-01   &~~ 7.59355110E-03     &~~ -1.0619821346E-01   \\
$b_{5	}$&~~ -3.8483814584E-01 &~~ 1.0539112890E-01    &~~ 9.25808533E-01     &~~ 2.5725126407E-01    \\
$b_{6	}$&~~ -1.7358035943E-01 &~~ -1.2348451952E-01   &~~ 2.40148437E-02     &~~ -9.0533064633E-02   \\
$b_{7	}$&~~ -1.0253728235E+00 &~~ -1.4515728482E+00   &~~ -1.18330413E+00    &~~ -1.6026688840E+00   \\
$b_{8	}$&~~ 2.1180011100E-01  &~~ 7.9827705222E-02    &~~ -4.83942869E-02    &~~ 2.5415111065E-02    \\
$b_{9	}$&~~ -3.2775987039E-01 &~~ -1.9621386113E-01   &~~ -1.33205930E-01    &~~ -1.3239995407E-01   \\
$b_{10}$&~~ 3.8408766601E+00  &~~ 1.9580717285E+00    &~~ 6.85987190E-01     &~~ 1.2754778849E+00    \\
$b_{11}$&~~ 1.8981173423E-01  &~~ 9.9713522137E-02    &~~ 5.66233393E-02     &~~ 5.7343326344E-02    \\
$b_{12}$&~~ -3.7600583268E-01 &~~ -2.1351883409E-01   &~~ -2.55948286E-01    &~~ -1.3703759581E-01   \\
$b_{13}$&~~ -5.4083155328E+00 &~~ -3.4649044585E-01   &~~ -5.40879277E-01    &~~ 4.2430865177E+00    \\
$b_{14}$&~~ 3.8805028093E+00  &~~ -8.6166033197E-01   &~~ -3.40328042E+00    &~~ -2.9485888052E+00   \\
$b_{15}$&~~ -1.8429751923E+00 &~~ -4.3744482206E-01   &~~ 1.44254957E+00     &~~ 5.4699391107E-02    \\
$b_{16}$&~~ 1.8302463857E+00  &~~ 9.9509140275E-01    &~~ 5.42811390E-01     &~~ 6.2922819182E-01    \\
$b_{17}$&~~ -7.2312391343E-01 &~~ -4.4774174549E-01   &~~ 3.76265531E-01     &~~ -3.0267386270E-01   \\

\hline \hline
\end{tabular}
\end{table}
\begin{table}[h!]
\centering 
\caption{The coefficients of $m \neq l$ modes which are valid for arbitrary spin parameters (continued from the previous Table).}

\label{coeflnm2}
\begin{tabular}{c| r r r r r}
\hline \hline
$ $&$l=4,m=2~~~~$&$l=5,m=4~~~~~$&$l=5,m=3~~~~$&$l=6,m=5~~~~~$&$l=6,m=4~~~~$
 \\

\hline
$a_1 $  & -8.67986098E-01 & -1.2781593391E+00  & -9.61379517E-01   & -1.2846696264E+00  & -1.02368654E+00 \\
$a_2 $  & -4.68123360E-01 & -9.7034478849E-01  & -6.73122187E-01   & -1.0280853586E+00  & -8.01814173E-01 \\
$a_3 $  & -5.32842812E+00 & 2.8215508168E+00   & 6.60464386E+00    & 1.4723471598E+00   & 1.33175794E+01  \\
$a_4 $  & -2.56745400E+01 & 1.2611970259E+01   & -2.30576646E+01   & 1.2336196550E+01   & -1.94946427E+01 \\
$a_5 $  & 5.93078462E+02  & -5.2596249376E+02  & 8.05350180E+02    & -6.2959344624E+02  & 8.25651691E+02  \\
$a_6 $  & -8.12541432E+02 & 3.0038340199E+02   & -7.08992989E+02   & 3.3177604825E+02   & -5.80599654E+02 \\
$a_7 $  & 6.30794899E+03  & -1.7661904159E+03  & 6.82178259E+03    & -2.5543377525E+03  & 6.36379605E+03  \\
$a_8 $  & -3.74091023E+03 & 6.0774411335E+02   & -3.26018905E+03   & 8.8256123155E+02   & -2.67275782E+03 \\
$a_9 $  & 5.93846520E+03  & -2.4280639567E+02  & 5.86575394E+03    & -1.0359578359E+03  & 5.20057093E+03  \\
$a_{10}$& -1.96409428E+03 & 3.2316814336E+01   & -1.71089135E+03   & 2.3941084370E+02   & -1.39883546E+03 \\
$b_{1	}$& 2.64746247E-05  & -2.4005464853E-01  & 8.63749612E-06    & -1.9451648765E-01  & -9.27611511E-06 \\
$b_{2	}$& 6.53413116E-06  & -1.1607843889E-01  & 1.22992270E-04    & -9.5157480501E-02  & 2.02754361E-04  \\
$b_{3	}$& -1.42108748E+00 & 3.2207333467E-01   & -1.18117099E+00   & 1.6169577622E-01   & -1.03742903E+00 \\
$b_{4	}$& 6.02315654E-03  & -8.2537773600E-02  & 5.70124410E-03    & -6.6960649581E-02  & 5.27509539E-03  \\
$b_{5	}$& 6.91692847E-01  & 3.3019097935E-01   & 5.82552838E-01    & 3.7135374972E-01   & 5.26944685E-01  \\
$b_{6	}$& 1.99698676E-02  & -7.1149022817E-02  & 1.69743369E-02    & -5.8801917957E-02  & 1.38468288E-02  \\
$b_{7	}$& -7.51628871E-01 & -1.6956674019E+00  & -7.46512958E-01   & -1.7616806546E+00  & -8.30305798E-01 \\
$b_{8	}$& -3.61239291E-02 & 3.9055128236E-03   & -5.03855988E-02   & -8.9134652221E-03  & -5.77294054E-02 \\
$b_{9	}$& -1.13130256E-01 & -9.7958187201E-02  & -9.31285992E-02   & -7.6563516673E-02  & -7.71485362E-02 \\
$b_{10}$& 4.37774719E-01  & 9.2758969640E-01   & 3.70643502E-01    & 7.2627287647E-01   & 3.36004880E-01  \\
$b_{11}$& 1.39342313E-02  & 4.1282519376E-02   & 3.80915154E-03    & 3.3347198540E-02   & -1.00856304E-03 \\
$b_{12}$& -1.83056410E-01 & -9.5252443489E-02  & -1.51139010E-01   & -6.8823516663E-02  & -1.23644570E-01 \\
$b_{13}$& 1.40858563E+00  & 8.0358698360E+00   & 4.00274644E+00    & 1.1235333168E+01   & 6.66317411E+00  \\
$b_{14}$& -2.75888278E+00 & -4.2196167836E+00  & -3.38216203E+00   & -5.1157622056E+00  & -4.04425061E+00 \\
$b_{15}$& 9.17504899E-01  & 2.7771091154E-01   & 8.19962357E-01    & 4.1191606722E-01   & 7.79402004E-01  \\
$b_{16}$& 6.25718075E-01  & 4.4088370234E-01   & 5.06908653E-01    & 3.2549939084E-01   & 4.11810202E-01  \\
$b_{17}$& 2.07386892E-01  & -2.2006697127E-01  & 1.76762132E-01    & -1.6838546731E-01  & 1.50191073E-01  \\

\hline \hline
\end{tabular}
\end{table}

\begin{table}[h!]
\centering 
\caption{The coefficients of $m \neq l$ modes which are valid for arbitrary spin parameters (continued from the previous Table).}
\label{coeflnm3}

\begin{tabular}{c| c c c c }
\hline \hline
$ $&$l=7,m=6$&$l=7,m=5$&$l=8,m=7$&$l=8,m=6$
 \\

\hline
$a_1 $  &~~-1.2868856926E+00 &~~ -1.06718606E+00 &~~ -1.2871564053E+00  &~~-1.09862959E+00 \\
$a_2 $  &~~-1.0742998386E+00 &~~ -8.72925981E-01 &~~ -1.1128440247E+00  &~~-9.49333228E-01 \\
$a_3 $  &~~3.5389121101E-01  &~~ 1.47083995E+01  &~~ -4.4832567658E-01  &~~1.68112921E+01  \\
$a_4 $  &~~1.2117041444E+01  &~~ -1.33425845E+01 &~~ 1.1779197256E+01   &~~-1.09584619E+01 \\
$a_5 $  &~~-7.2029338524E+02 &~~ 6.44931928E+02  &~~ -7.9134006497E+02  &~~5.13352447E+02  \\
$a_6 $  &~~3.5430426346E+02  &~~ -4.00244363E+02 &~~ 3.6700960186E+02   &~~-2.81718147E+02 \\
$a_7 $  &~~-3.2214079577E+03 &~~ 5.01794687E+03  &~~ -3.7431050034E+03  &~~3.63541010E+03  \\
$a_8 $  &~~1.0805133404E+03  &~~ -1.94007255E+03 &~~ 1.2133023551E+03   &~~-1.30245487E+03 \\
$a_9 $  &~~-1.6725256262E+03 &~~ 4.09493264E+03  &~~ -2.1631469998E+03  &~~2.76962485E+03  \\
$a_{10}$&~~3.8847848728E+02  &~~ -1.04137123E+03 &~~ 4.9308096909E+02   &~~-6.69785919E+02 \\
$b_{1	}$&~~-1.6334989829E-01 &~~ -2.03849855E-05 &~~ -1.4071907195E-01  &~~-3.84478649E-05 \\
$b_{2	}$&~~-8.0420453056E-02 &~~ 2.42955919E-04  &~~ -6.9527455767E-02  &~~2.94754750E-04  \\
$b_{3	}$&~~4.6726322034E-02  &~~ -9.45261092E-01 &~~ -3.9737638808E-02  &~~-8.82296964E-01 \\
$b_{4	}$&~~-5.6038220422E-02 &~~ 4.44161354E-03  &~~ -4.7985813649E-02  &~~4.47231263E-03  \\
$b_{5	}$&~~3.9705858531E-01  &~~ 4.97496977E-01  &~~ 4.1426461301E-01   &~~4.79546435E-01  \\
$b_{6	}$&~~-5.0353124839E-02 &~~ 9.85139353E-03  &~~ -4.4237741136E-02  &~~8.13447434E-03  \\
$b_{7	}$&~~-1.8115776972E+00 &~~ -9.22263692E-01 &~~ -1.8510337188E+00  &~~-1.02566083E+00 \\
$b_{8	}$&~~-1.7980166648E-02 &~~ -5.32894531E-02 &~~ -2.5147833232E-02  &~~-6.24697595E-02 \\
$b_{9	}$&~~-6.2211796357E-02 &~~ -6.60489280E-02 &~~ -5.1982278116E-02  &~~-5.51204324E-02 \\
$b_{10}$&~~5.9892343420E-01  &~~ 3.02149774E-01  &~~ 5.1314123950E-01   &~~2.82743530E-01  \\
$b_{11}$&~~2.8581418396E-02  &~~ 7.36668700E-03  &~~ 2.5186758681E-02   &~~-1.14591459E-03 \\
$b_{12}$&~~-5.0830612068E-02 &~~ -1.02138072E-01 &~~ -3.7884556226E-02  &~~-8.18636393E-02 \\
$b_{13}$&~~1.3966273925E+01  &~~ 8.96591811E+00  &~~ 1.6333115342E+01   &~~1.14813438E+01  \\
$b_{14}$&~~-5.7857929460E+00 &~~ -4.54168077E+00 &~~ -6.3083173761E+00  &~~-5.12636320E+00 \\
$b_{15}$&~~5.0357745315E-01  &~~ 7.30217085E-01  &~~ 5.7163615508E-01   &~~7.58568387E-01  \\
$b_{16}$&~~2.4881708383E-01  &~~ 3.42234873E-01  &~~ 1.9460946007E-01   &~~2.75045222E-01  \\
$b_{17}$&~~-1.3380632914E-01 &~~ 1.28036706E-01  &~~ -1.0936772811E-01  &~~1.11398836E-01  \\

\hline \hline
\end{tabular}
\end{table}

To demonstrate the performances of these recalibration formalisms, we show the relative differences between our recalibrated fluxes and the Teukolsky-based ones in Fig. \ref{ge8}.
\begin{figure}
\begin{center}
\includegraphics[height=3in]{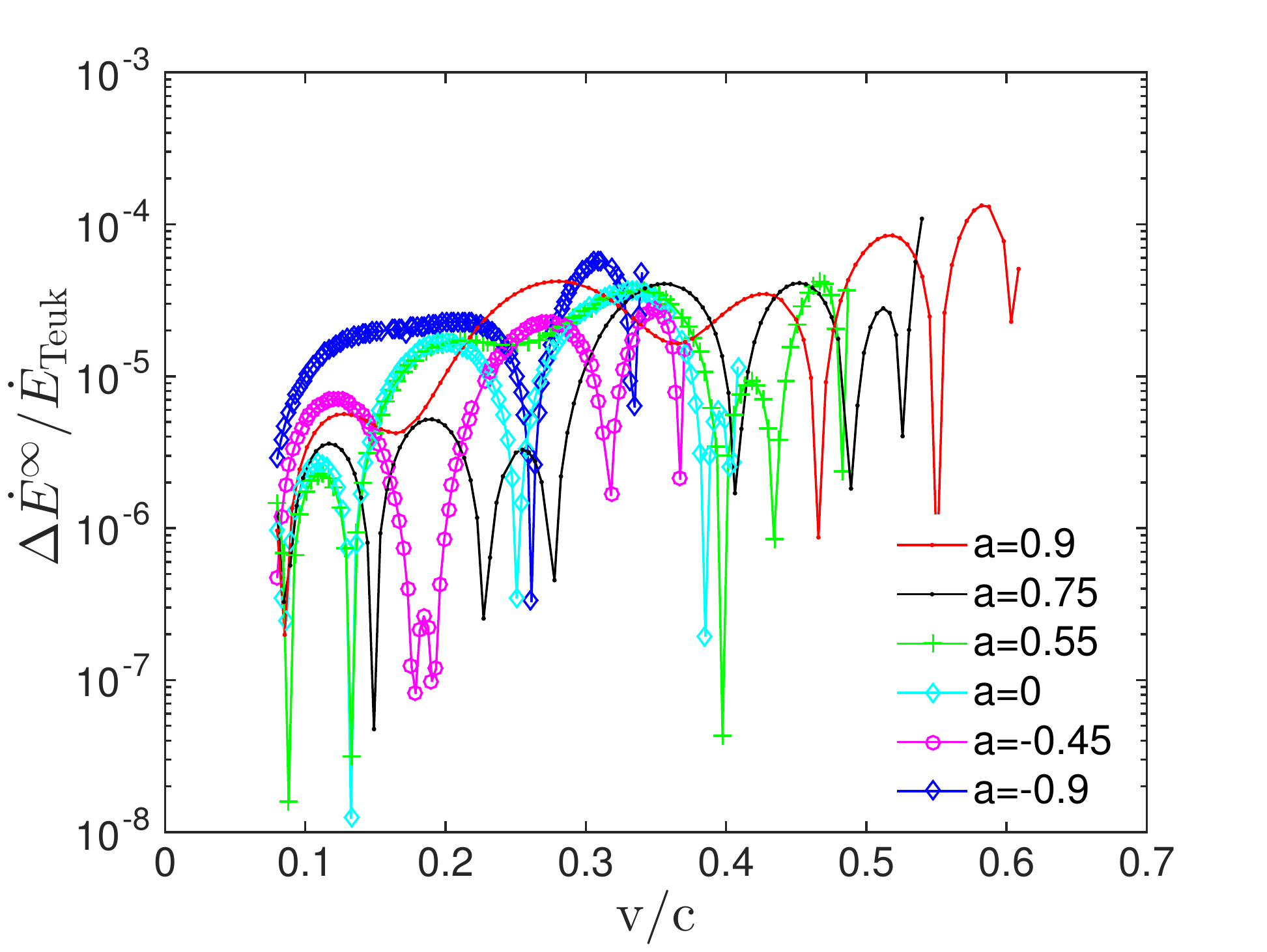}
\caption{The relative errors of the energy fluxes emission to infinity. $\Delta \dot{E}^\infty \equiv |\dot{E}^\infty_{\rm Teuk} - \dot{E}^\infty_{\rm recal}|$, where $\dot{E}^\infty_{\rm recal}$ means the recalibrated PN fluxes. } \label{ge8}
\end{center}
\end{figure}
From Fig. \ref{ge8} we can find that the maximum relative error of our recalibrated fluxes is just a little larger than $1\times10^{-4}$. Comparing with the Fig. 1 in Ref. \cite{Yunes11}, we can find that our new formalisms are not only much better than the original factorized-resummation PN fluxes (relative errors are even larger than $10^{-1}$), but also a little better than the calibrated results of Yunes et al (the maximum relative error approaches $10^{-3}$).

However, based on the experiences in Ref. \cite{Yunes11}, only one order of improvement may be not enough for the EMRI simulations with long time scale. We find that it is very hard to get the more accurate ``global" coefficients which are valid for all $q$ parameters. For getting the better performance, the spin background is divided into 4 groups(see Table \ref{qgroup}) under consideration, and for each $q$ group, coefficients $a_n$ and $b_n$ in Eqs. (\ref{le2m}) are fitted by the least squares method for this group only. These grouped coefficients are listed in Table \ref{lcoef1}-\ref{lcoef4}. Note that we still use the ``global" coefficients for the $m \neq l$ cases in Eq. (\ref{lne2m}). This is because $l=m$ modes are much more dominant than $l \neq m$ ones for the same $l$, then the accuracy of fluxes is mainly determined by the $l=m$ modes. For simplicity, in the cases of $l \neq m$ we still use the ``global" coefficients listed in Tables \ref{coeflnm1}-\ref{coeflnm3} for all $q$ values. For some unimportant modes (such as l=4, m=1), we directly use the original PN formalisms which can be find in Ref. \cite{panyi11}. 
\begin{table}[h!]
\caption{Groups of $q$ ($m = l$)}
\label{qgroup}
\begin{tabular}{c| c  }
\hline \hline

group&$q$\\
 \hline
I&-0.9 $\sim$ -0.3\\
II&-0.3 $\sim$ 0.3\\
III&0.3 $\sim$ 0.7\\
IV&0.7 $\sim$ 0.9\\

\hline
\hline \hline
\end{tabular}
\end{table}

\begin{figure}
\begin{center}
\includegraphics[height=3in]{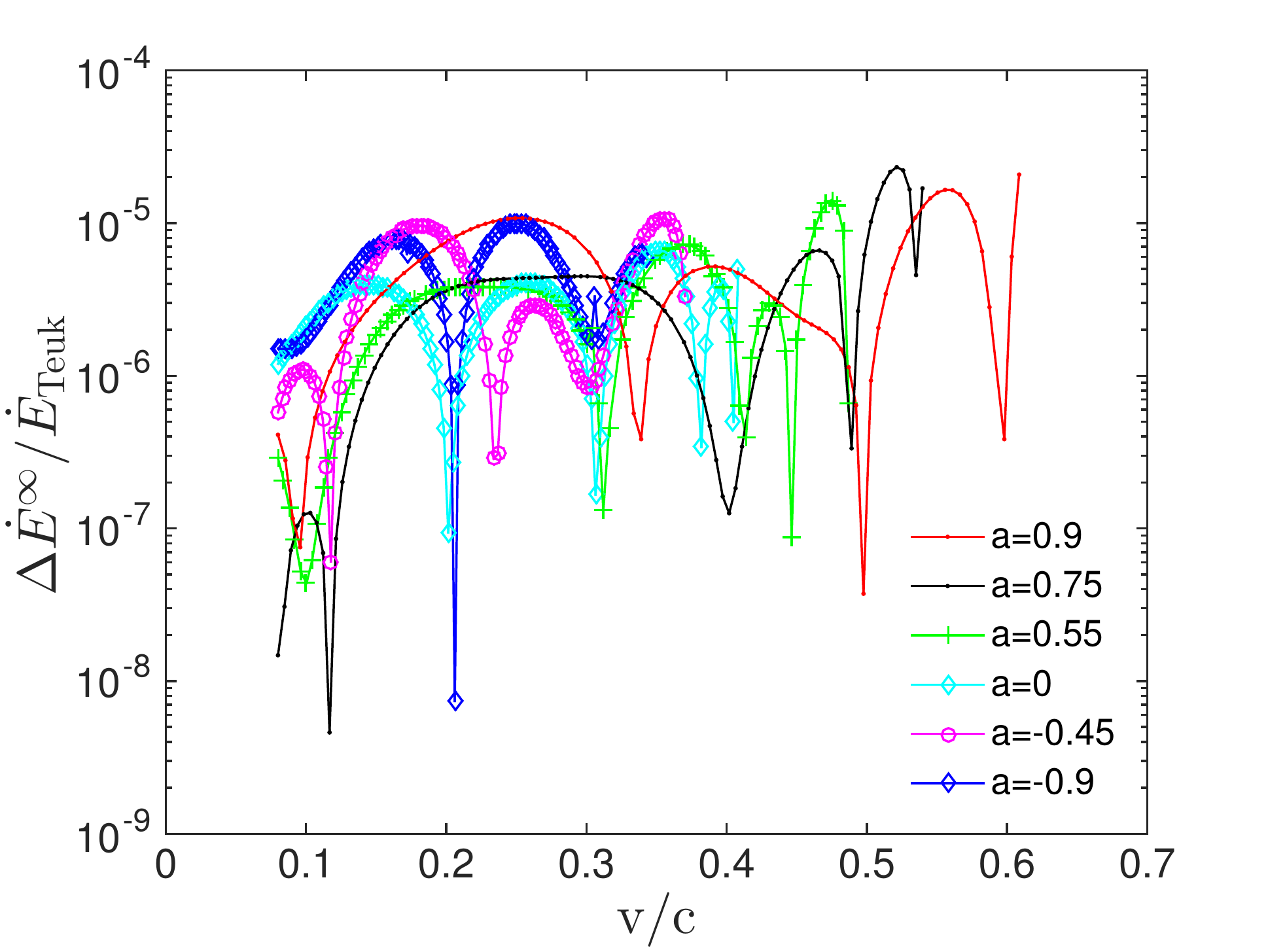}
\caption{The relative errors of the grouped-recalibration energy fluxes emission to infinity.} \label{le8}
\end{center}
\end{figure}

The results of recalibrated fluxes using the grouped coefficients in Tables \ref{lcoef1}-\ref{lcoef4} are shown in Fig. \ref{le8}. We can find now that the relative precision is improved one order compared with the ``global" ones. The maximum relative error is just a little larger than $1\times 10^{-5}$. We can conclude that our recalibrated waveforms are more precise than the ones in Ref. \cite{Yunes11} for about two orders.

\begin{sidewaystable}
\centering
\begin{threeparttable}[b]
\caption{The coefficients of $m = l$ modes which are valid for the cases of $q\in [-0.9,-0.3]$.}
\label{lcoef1}
\footnotesize
\begin{tabular}{c| c c c c c c c }
\hline \hline
\hspace{-1in}$ $&$l=2,m=2$&$l=3,m=3$&$l=4,m=4$&$l=5,m=5$&$l=6,m=6$&$l=7,m=7$&$l=8,m=8$
 \\

\hline
$a_1$   &~ -1.023832458915E+00	&~ -1.166670930290E+00	&~ -1.2227269230E+00	&~ -1.2487112825E+00	&~ -1.2618970998E+00	&~ -1.2688994648E+00	&~ -1.2726525633E+00  \\
$a_2$   &~ -1.932219959525E+00	&~ -1.698223965937E+00	&~ -1.6090214076E+00	&~ -1.5763749717E+00	&~ -1.5557234483E+00	&~ -1.5432721072E+00	&~ -1.5351556938E+00  \\
$a_3$   &~ 1.677638932194E+01	&~ 1.352289231106E+01	&~ 1.5234398474E+01	&~ 1.5939741413E+01	&~ 1.6846966257E+01	&~ 1.7626876023E+01	&~ 1.8311555540E+01   \\
$a_4$   &~ 1.547950099412E+00	&~ -4.594744372273E+00	&~ -1.5336197229E+00	&~ -4.6068208501E+00	&~ -4.6321041996E+00	&~ -4.6549632411E+00	&~ -4.6789608178E+00  \\
$a_5$	  &~ -1.819262450631E+02	&~ 4.449131933181E+01	&~ -1.6348633036E+02	&~ 9.6188943435E+01	&~ 1.1286127182E+02	&~ 1.2744095189E+02	&~ 1.4114127162E+02   \\
$a_6$	  &~ 3.219466389529E+02	&~ -1.172363806204E+02	&~ 1.4985145089E+02	&~ -8.8110830773E+01	&~ -8.6709862977E+01	&~ -8.6870332164E+01	&~ -8.7986688726E+01  \\
$a_7$	  &~ -2.827834157899E+03	&~ 3.146030860193E+03	&~ -2.7493929411E+03	&~ 2.3775508391E+03	&~ 2.4179488592E+03	&~ 2.5289425461E+03	&~ 2.6827420048E+03   \\
$a_8$	  &~ 1.159839574546E+03	&~ -2.218252260934E+03	&~ 1.3314501667E+03	&~ -1.1348471615E+03	&~ -1.0455018964E+03	&~ -1.0187876232E+03	&~ -1.0239783955E+03  \\
$a_9$	  &~ 3.860142871343E+03	&~ 1.181743240069E+04	&~ -4.4091821914E+03	&~ 5.5928283348E+03	&~ 5.1034145804E+03	&~ 5.0398461467E+03	&~ 5.2045393560E+03   \\
$a_{10}$&~ -2.196778064836E+03	&~ -4.177397713728E+03	&~ 1.2994684071E+03	&~ -1.6151861628E+03	&~ -1.3860241588E+03	&~ -1.3077178274E+03	&~ -1.3037896868E+03  \\
$b_{1 }$&~ -6.679322945322E-01	&~ -6.672042477171E-01	&~ -6.6677424197E-01	&~ -6.6664482663E-01	&~ -6.6659025716E-01	&~ -6.6656792342E-01	&~ -6.6656133393E-01  \\
$b_{2 }$&~ 4.823667901085E-01	&~ 4.936606033319E-01	&~ 4.9842595539E-01	&~ 4.9973876152E-01	&~ 5.0031960298E-01	&~ 5.0053578364E-01	&~ 5.0057606118E-01   \\
$b_{3 }$&~ -1.602055927422E+00	&~ -1.288152428516E+00	&~ -1.2173825540E+00	&~ -1.2071130046E+00	&~ -1.2157302760E+00	&~ -1.2305441193E+00	&~ -1.2468036358E+00  \\
$b_{4 }$&~ 6.174319215479E-01	&~ 2.302527921765E-01	&~ 5.1681459427E-02	&~ 1.6268466934E-03	&~ -2.5498005329E-02	&~ -3.7363543830E-02	&~ -4.1716490850E-02  \\
$b_{5 }$&~ 1.878095021311E+00	&~ 1.773670647475E+00	&~ 1.9466888064E+00	&~ 2.0894528027E+00	&~ 2.1465897448E+00	&~ 2.1803529023E+00	&~ 2.2013034508E+00   \\
$b_{6 }$&~ -1.755010624878E+00	&~ -3.501376465865E-01	&~ 2.2268398735E-01	&~ 4.8501154413E-01	&~ 5.6907258126E-01	&~ 6.0228737015E-01	&~ 6.1093515077E-01   \\
$b_{7 }$&~ -1.088444589142E+01	&~ -4.463661653572E+00	&~ -7.6660773973E-01	&~ 6.3709138130E-01	&~ 1.1794016785E+00	&~ 1.4121406028E+00	&~ 1.4931722527E+00   \\
$b_{8 }$&~ -5.789635190798E+00	&~ -2.036898753679E+00	&~ -3.6653871468E-01	&~ 3.2366438920E-01	&~ 5.6178386987E-01	&~ 6.5607418337E-01	&~ 6.8094253934E-01   \\
$b_{9 }$&~ -3.047757117233E+01	&~ -1.434987580728E+01	&~ -9.9155001811E+00	&~ -8.6917434771E+00	&~ -8.2775468846E+00	&~ -8.2201526148E+00	&~ -8.3255558909E+00  \\
$b_{10}$&~ -2.095203739841E+01	&~ -6.692674618331E+00	&~ 3.4183209832E-01	&~ 2.9961775728E+00	&~ 3.9657245653E+00	&~ 4.3452919347E+00	&~ 4.4412052262E+00   \\

\hline \hline
\end{tabular}

  \begin{tablenotes}
    \item [a] { }
   \end{tablenotes}
\end{threeparttable}
\end{sidewaystable}
\clearpage

\begin{sidewaystable}
\centering
\begin{threeparttable}[b]
\caption{The coefficients of $m = l$ modes which are valid for the cases of $q\in [-0.3,~0.3]$.}
\label{lcoef2}
\footnotesize
\begin{tabular}{c| c c c c c c c }
\hline \hline
\hspace{-1in}$ $&$l=2,m=2$&$l=3,m=3$&$l=4,m=4$&$l=5,m=5$&$l=6,m=6$&$l=7,m=7$&$l=8,m=8$
 \\

\hline
$a_1$   &~ -1.023911645424E+00	&~ -1.166704782425E+00	&~ -1.2227503690E+00	&~ -1.2487360657E+00	&~ -1.2619211944E+00	&~ -1.2689236287E+00	&~ -1.2726771427E+00  \\
$a_2$   &~ -1.863038361493E+00	&~ -1.668391076204E+00	&~ -1.5966889000E+00	&~ -1.5625464596E+00	&~ -1.5433377230E+00	&~ -1.5312744657E+00	&~ -1.5230412385E+00  \\
$a_3$   &~ 3.322718074004E+01	&~ 1.665960902214E+01	&~ 1.4695046047E+01	&~ 1.4556814927E+01	&~ 1.4694384382E+01	&~ 1.4798293040E+01	&~ 1.4808780845E+01   \\
$a_4$   &~ 2.730630887841E+01	&~ 7.224840921117E+00	&~ 2.5830515593E+00	&~ 9.2394293900E-01	&~ 3.1986154779E-01	&~ 1.3940390016E-01	&~ 1.6042892981E-01   \\
$a_5$	  &~ -8.304618864711E+02	&~ -5.422406825013E+02	&~ -4.0614961025E+02	&~ -3.4709345755E+02	&~ -3.3523568079E+02	&~ -3.4836290526E+02	&~ -3.7633406435E+02  \\
$a_6$	  &~ 1.794927232286E+03	&~ 6.091326360516E+02	&~ 3.3903543143E+02	&~ 2.3977971693E+02	&~ 2.0349336799E+02	&~ 1.9251520258E+02	&~ 1.9355761019E+02   \\
$a_7$	  &~ -1.986790361983E+04	&~ -8.396711985052E+03	&~ -5.2514311123E+03	&~ -3.9302101508E+03	&~ -3.4830399399E+03	&~ -3.4397538487E+03	&~ -3.6147613721E+03  \\
$a_8$	  &~ 1.389486446504E+04	&~ 4.513748998687E+03	&~ 2.4212312592E+03	&~ 1.6202222857E+03	&~ 1.3230758432E+03	&~ 1.2291277108E+03	&~ 1.2320204768E+03   \\
$a_9$	  &~ -3.360589053649E+04	&~ -1.193004353708E+04	&~ -6.6891839081E+03	&~ -4.4255626631E+03	&~ -3.5776443478E+03	&~ -3.3522278626E+03	&~ -3.4499340346E+03  \\
$a_{10}$&~ 1.213431060551E+04	&~ 3.720835603371E+03	&~ 1.9006928587E+03	&~ 1.1667089974E+03	&~ 8.9032833062E+02	&~ 7.9903937028E+02	&~ 7.9626881418E+02   \\
$b_{1 }$&~ -6.625553460827E-01	&~ -6.651912993081E-01	&~ -6.6583126754E-01	&~ -6.6602585321E-01	&~ -6.6609082163E-01	&~ -6.6610350592E-01	&~ -6.6609178558E-01  \\
$b_{2 }$&~ 5.084738194434E-01	&~ 5.018306197183E-01	&~ 4.9890829771E-01	&~ 4.9790333490E-01	&~ 4.9745030177E-01	&~ 4.9723071286E-01	&~ 4.9712922739E-01   \\
$b_{3 }$&~ -1.759096174739E+00	&~ -1.349399188167E+00	&~ -1.2492538262E+00	&~ -1.2293579576E+00	&~ -1.2347072456E+00	&~ -1.2486721523E+00	&~ -1.2652216878E+00  \\
$b_{4 }$&~ -4.228398000760E-02	&~ 5.589578425315E-03	&~ 1.9329158666E-02	&~ 1.6547069593E-02	&~ 1.3835523345E-02	&~ 1.1880704141E-02	&~ 1.0361359195E-02   \\
$b_{5 }$&~ 3.800658995049E+00	&~ 2.511213247521E+00	&~ 2.2840343352E+00	&~ 2.2482977164E+00	&~ 2.2535960502E+00	&~ 2.2712850380E+00	&~ 2.2930950352E+00   \\
$b_{6 }$&~ 7.291799468567E-02	&~ 2.475215605027E-01	&~ 2.8514659642E-01	&~ 2.9680305212E-01	&~ 3.0234919450E-01	&~ 3.0513481524E-01	&~ 3.0650489157E-01   \\
$b_{7 }$&~ 4.502358337784E+00	&~ 8.622007126069E-01	&~ 1.7623366766E-03	&~ -2.5771960742E-01	&~ -3.4282462355E-01	&~ -3.5866130083E-01	&~ -3.4315549387E-01  \\
$b_{8 }$&~ -2.076565859479E-01	&~ -9.117687595244E-02	&~ -9.7885048071E-02	&~ -9.5954391716E-02	&~ -9.4412344504E-02	&~ -9.3277582742E-02	&~ -9.2367757694E-02  \\
$b_{9 }$&~ -2.642956467487E+01	&~ -1.317811513496E+01	&~ -1.0491249891E+01	&~ -9.7523765390E+00	&~ -9.5822087233E+00	&~ -9.6344901034E+00	&~ -9.7836542572E+00  \\
$b_{10}$&~ 2.832079595588E+00	&~ 1.735525732773E+00	&~ 1.5962882558E+00	&~ 1.5480773313E+00	&~ 1.5175693924E+00	&~ 1.4972347754E+00	&~ 1.4830004367E+00   \\

\hline \hline
\end{tabular}

  \begin{tablenotes}
    \item [a] { }
   \end{tablenotes}
\end{threeparttable}
\end{sidewaystable}
\clearpage

\begin{sidewaystable}
\centering
\begin{threeparttable}[b]
\caption{The coefficients of $m = l$ modes which are valid for the cases of $q\in [0.3, ~0.7]$.}
\label{lcoef3}
\footnotesize
\begin{tabular}{c| c c c c c c c }
\hline \hline
\hspace{-1in}$ $&$l=2,m=2$&$l=3,m=3$&$l=4,m=4$&$l=5,m=5$&$l=6,m=6$&$l=7,m=7$&$l=8,m=8$
 \\

\hline
$a_1$   &~ -1.023975805956E+00	&~ -1.166721079966E+00	&~ -1.2227931618E+00	&~ -1.2487811477E+00	&~ -1.2619660215E+00	&~ -1.2689701639E+00	&~ -1.2727256099E+00  \\
$a_2$   &~ -1.773785209399E+00	&~ -1.632220902128E+00	&~ -1.5488861713E+00	&~ -1.5137494102E+00	&~ -1.4927232619E+00	&~ -1.4786712228E+00	&~ -1.4685006871E+00  \\
$a_3$   &~ 3.318171742361E+01	&~ 1.485007940230E+01	&~ 1.1208215801E+01	&~ 8.7424413748E+00	&~ 6.5654511682E+00	&~ 4.4639819195E+00	&~ 2.4007004823E+00   \\
$a_4$   &~ 3.953835298796E+01	&~ 1.245654119994E+01	&~ 1.2709894830E+01	&~ 1.1883585053E+01	&~ 1.1939038028E+01	&~ 1.2309585938E+01	&~ 1.2798559744E+01   \\
$a_5$	  &~ -1.104050524831E+03	&~ -6.678847665787E+02	&~ -8.6428660908E+02	&~ -9.5407467730E+02	&~ -1.0672751384E+03	&~ -1.1881289093E+03	&~ -1.3101960825E+03  \\
$a_6$	  &~ 1.765670955233E+03	&~ 6.364340267479E+02	&~ 6.4805577773E+02	&~ 6.1182473525E+02	&~ 6.1235129084E+02	&~ 6.2613175012E+02	&~ 6.4502502325E+02   \\
$a_7$	  &~ -1.584521768582E+04	&~ -7.018461656809E+03	&~ -8.2330639578E+03	&~ -8.5105354171E+03	&~ -9.1262632809E+03	&~ -9.8614596889E+03	&~ -1.0634914259E+04  \\
$a_8$	  &~ 1.007746475744E+04	&~ 3.520695972919E+03	&~ 3.6160554839E+03	&~ 3.4023945061E+03	&~ 3.4000165921E+03	&~ 3.4742319939E+03	&~ 3.5784770733E+03   \\
$a_9$	  &~ -1.943757060254E+04	&~ -7.516377182961E+03	&~ -8.4638217440E+03	&~ -8.4396976216E+03	&~ -8.8310465543E+03	&~ -9.3720349574E+03	&~ -9.9671709023E+03  \\
$a_{10}$&~ 6.701134943008E+03	&~ 2.262034338117E+03	&~ 2.3482970193E+03	&~ 2.2049397140E+03	&~ 2.2024481099E+03	&~ 2.2510887696E+03	&~ 2.3199082137E+03   \\
$b_{1 }$&~ -6.733755278411E-01	&~ -6.696972546143E-01	&~ -6.6915298811E-01	&~ -6.6901906546E-01	&~ -6.6904049613E-01	&~ -6.6911153302E-01	&~ -6.6919077151E-01  \\
$b_{2 }$&~ 5.843602324003E-01	&~ 5.360214944791E-01	&~ 5.2813489804E-01	&~ 5.2594458075E-01	&~ 5.2571830732E-01	&~ 5.2623809885E-01	&~ 5.2706490173E-01   \\
$b_{3 }$&~ -1.662151538548E+00	&~ -1.308471947237E+00	&~ -1.2252358713E+00	&~ -1.2090008750E+00	&~ -1.2146809603E+00	&~ -1.2277311411E+00	&~ -1.2428755772E+00  \\
$b_{4 }$&~ -1.061146392648E+00	&~ -4.975131465872E-01	&~ -4.5878852649E-01	&~ -4.6545217148E-01	&~ -4.8516130249E-01	&~ -5.0792509318E-01	&~ -5.3091243278E-01  \\
$b_{5 }$&~ 3.827859278692E+00	&~ 2.543651584540E+00	&~ 2.4583522998E+00	&~ 2.4535277945E+00	&~ 2.4730693730E+00	&~ 2.4981097844E+00	&~ 2.5238213771E+00   \\
$b_{6 }$&~ -7.756293634998E-01	&~ -3.514346900192E-01	&~ -1.8135025452E-01	&~ -1.4988716272E-01	&~ -1.4384902166E-01	&~ -1.4725559217E-01	&~ -1.5457211966E-01  \\
$b_{7 }$&~ 1.045517070565E+01	&~ 4.621358937544E+00	&~ 3.5130160665E+00	&~ 3.3478120875E+00	&~ 3.4184709464E+00	&~ 3.5704120877E+00	&~ 3.7502614404E+00   \\
$b_{8 }$&~ 1.320082877129E+00	&~ 1.248327500336E+00	&~ 1.0815767083E+00	&~ 1.1204618339E+00	&~ 1.1773606154E+00	&~ 1.2360870006E+00	&~ 1.2922880604E+00   \\
$b_{9 }$&~ -2.960658687909E+01	&~ -1.530065544785E+01	&~ -1.2893121094E+01	&~ -1.2295951469E+01	&~ -1.2251911718E+01	&~ -1.2416517301E+01	&~ -1.2665914817E+01  \\
$b_{10}$&~ -7.235679546357E-01	&~ -1.490112966956E+00	&~ -1.5024171205E+00	&~ -1.7897701891E+00	&~ -2.0560884742E+00	&~ -2.2948708721E+00	&~ -2.5082509405E+00  \\

\hline \hline
\end{tabular}

  \begin{tablenotes}
    \item [a] { }
   \end{tablenotes}
\end{threeparttable}
\end{sidewaystable}
\clearpage

\begin{sidewaystable}
\centering
\begin{threeparttable}[b]
\caption{The coefficients of $m = l$ modes which are valid for the cases of $q\in [0.7,~0.9]$.}
\label{lcoef4}
\footnotesize
\begin{tabular}{c| c c c c c c c }
\hline \hline
\hspace{-1in}$ $&$l=2,m=2$&$l=3,m=3$&$l=4,m=4$&$l=5,m=5$&$l=6,m=6$&$l=7,m=7$&$l=8,m=8$
 \\

\hline
$a_1$   &~ -1.023713029818E+00	&~ -1.166658087088E+00	&~ -1.2227316403E+00	&~ -1.2487204975E+00	&~ -1.2619126669E+00	&~ -1.2689160680E+00	&~ -1.2726695812E+00 \\
$a_2$   &~ -1.794624263575E+00	&~ -1.621193222378E+00	&~ -1.5327448233E+00	&~ -1.4942946985E+00	&~ -1.4678080567E+00	&~ -1.4503622956E+00	&~ -1.4372711854E+00 \\
$a_3$   &~ 2.046890798294E+01	&~ 1.390712770516E+01	&~ 1.1567575503E+01	&~ 1.0194394264E+01	&~ 8.5289500625E+00	&~ 7.0033350488E+00	&~ 5.4770757367E+00  \\
$a_4$   &~ 1.622959908299E+01	&~ 5.362947561996E+00	&~ 5.9479551176E+00	&~ 5.7073781701E+00	&~ 6.3034863153E+00	&~ 6.7994477812E+00	&~ 7.3138819873E+00  \\
$a_5$	  &~ -5.004421386961E+02	&~ -3.347118008786E+02	&~ -4.1312857910E+02	&~ -4.4857048093E+02	&~ -5.1491441843E+02	&~ -5.7583770709E+02	&~ -6.3683030983E+02 \\
$a_6$	  &~ 6.231165980424E+02	&~ 2.727905322411E+02	&~ 2.7099582189E+02	&~ 2.5560133274E+02	&~ 2.6511455316E+02	&~ 2.7437058032E+02	&~ 2.8509973772E+02  \\
$a_7$	  &~ -4.325486276431E+03	&~ -2.273153589664E+03	&~ -2.5435242101E+03	&~ -2.5976735403E+03	&~ -2.8623678749E+03	&~ -3.1141781636E+03	&~ -3.3740112438E+03 \\
$a_8$	  &~ 2.468012372878E+03	&~ 1.040409109822E+03	&~ 1.0288179150E+03	&~ 9.6282315003E+02	&~ 9.9350009082E+02	&~ 1.0260869242E+03	&~ 1.0649950447E+03  \\
$a_9$	  &~ -3.485493596189E+03	&~ -1.544287152302E+03	&~ -1.6616069282E+03	&~ -1.6397241447E+03	&~ -1.7728189304E+03	&~ -1.9041645860E+03	&~ -2.0434099896E+03 \\
$a_{10}$&~ 1.125528084941E+03	&~ 4.333129781275E+02	&~ 4.3301270223E+02	&~ 4.0415970515E+02	&~ 4.1884205004E+02	&~ 4.3461406272E+02	&~ 4.5305827253E+02  \\
$b_{1 }$&~ -6.797912744755E-01	&~ -6.720621720179E-01	&~ -6.7194729277E-01	&~ -6.7184933791E-01	&~ -6.7203482256E-01	&~ -6.7227724535E-01	&~ -6.7254064567E-01 \\
$b_{2 }$&~ 6.245384453404E-01	&~ 5.521354590036E-01	&~ 5.4154117537E-01	&~ 5.3623537128E-01	&~ 5.3423839936E-01	&~ 5.3351635286E-01	&~ 5.3344326317E-01  \\
$b_{3 }$&~ -2.475298992143E+00	&~ -1.873194636033E+00	&~ -1.7053905206E+00	&~ -1.6661363134E+00	&~ -1.6682748026E+00	&~ -1.6862288889E+00	&~ -1.7095463658E+00 \\
$b_{4 }$&~ 1.063403908967E+00	&~ 1.979531940044E+00	&~ 2.0150680610E+00	&~ 2.1826864324E+00	&~ 2.3500031062E+00	&~ 2.5103006064E+00	&~ 2.6568734489E+00  \\
$b_{5 }$&~ 7.422850698915E+00	&~ 1.859844594559E+00	&~ 7.1607997188E-01	&~ -7.7878036698E-02	&~ -6.4882027779E-01	&~ -1.1011487696E+00	&~ -1.4738948390E+00 \\
$b_{6 }$&~ -2.387763518986E+00	&~ -5.235497761725E+00	&~ -5.7336689248E+00	&~ -6.3980771439E+00	&~ -7.0116809124E+00	&~ -7.5697165115E+00	&~ -8.0687409701E+00 \\
$b_{7 }$&~ -6.881562105846E-01	&~ 6.911189729988E+00	&~ 9.0833201737E+00	&~ 1.1038471625E+01	&~ 1.2677125745E+01	&~ 1.4082846182E+01	&~ 1.5303204212E+01  \\
$b_{8 }$&~ -9.954888325283E-01	&~ 3.358493951171E+00	&~ 4.2258844731E+00	&~ 4.9786548018E+00	&~ 5.6027555463E+00	&~ 6.1391865950E+00	&~ 6.6040122375E+00  \\
$b_{9 }$&~ -3.036495432036E+01	&~ -1.795232730911E+01	&~ -1.5886923395E+01	&~ -1.5479532079E+01	&~ -1.5606259900E+01	&~ -1.5918834614E+01	&~ -1.6298827932E+01 \\
$b_{10}$&~ 1.372472520045E+01	&~ 2.847195462527E-01	&~ -2.9180442850E+00	&~ -5.0025500993E+00	&~ -6.5179130980E+00	&~ -7.7170426258E+00	&~ -8.7052997962E+00 \\

\hline \hline
\end{tabular}

  \begin{tablenotes}
    \item [a] { }
   \end{tablenotes}
\end{threeparttable}
\end{sidewaystable}
\clearpage
As we can see in reference \cite{Yunes10}, the PN formalisms for the energy fluxes absorbed by the event horizon perform quite inaccurate, so $\dot E^{\text H}$ is also supposed to be calibrated as above according to eq.(\ref{eh}) and the relevant explanation in our paper.
To improve the performance in estimating the energy flux, $a_8^{\rm Hor,NS}, a_{10}^{\rm Hor,NS}$ and $a_{12}^{\rm Hor,NS}$ are firstly selected to be refitted to replace the original PN parameters correspondingly for $q=0$ with 18 data by using the least-square method while staying other parameters unchanged(for $a^{\text{Hor,NS}}_{<8}=a^{\text{Hor,NS}}_{9}=a^{\text{Hor,NS}}_{11}=a^{\text{Hor,NS}}_{13}=0$), then under this precondition we consider the parameter-fitting in $a_{11}^{\rm Hor,S}$ and $a_{12}^{\rm Hor,S}$ for $q=$-0.9, -0.7, -0.5, -0.3, -0.1, 0, 0.1, 0.2, 0.3, 0.4, 0.5, 0.6, 0.7, 0.8 and 0.9, each with 18 data. And for simplicity, we adopt the following expressions for both terms and make $p_n$ fitted, 
\begin{align}
a_{11}^{\rm Hor,S}=&p_1q+p_2q^2+p_3q^3+p_4q^4+p_5q^5+p_6q^6  \nonumber\\
a_{12}^{\rm Hor,S}=&p_7q+p_8q^2+p_9q^3+p_{10}q^4+p_{11}q^5+p_{12}q^6+p_{13}q^7,
\end{align}
 and the results are listed in Table X. 

The comparison of the total PN fluxes with Teukolsky-based one is shown in Fig. \ref{ge8}. We can find that the recalibrated formalisms for the energy fluxes down to the horizon are much more accurate than the ones with original coefficients. Compared with the Fig. 2 of \cite{Yunes11}, the accuracy of the total energy fluxes is improved about two orders.

\begin{table}[h!]
\caption{Fitting results of $a_8^{\rm Hor,NS}, a_{10}^{\rm Hor,NS}, a_{12}^{\rm Hor,NS}$ and coefficients for $a_{11}^{\rm Hor,S}$ and $a_{12}^{\rm Hor,S}$}\vspace{0.1in}

\label{ghcoef}
\hspace{0.0in}

\footnotesize
\begin{tabular}{c c| c c  }
\hline \hline

$a_8^{\rm Hor,NS}$ &~~5.2372530107E+00~~
 &~~ $p_6$ &~~ 3.5515549857E+02~~
 \\
$a_{10}^{\rm Hor,NS}$ &~~-7.7738921771E+01~~
 &~~ $p_7$ &~~ -1.5904516176E+03~~
 \\
$a_{12}^{\rm Hor,NS}$ &~~4.5115894278E+02~~
 &~~ $p_8$ &~~ 1.5764986553E+03~~
 \\
$p_1$ &~~3.7707905550E+02~~
 &~~ $p_9$ &~~ -9.2841853361E+02~~
\\
$p_2$ &~~-2.6962149194E+02~~
 &~~ $p_{10}$ &~~ 1.2917574980E+03~~
\\
$p_3$ &~~8.9976203355E+01~~
 &~~ $p_{11}$ &~~ -2.0917886530E+02~~
\\
$p_4$ &~~-3.8417467857E+02~~
 &~~ $p_{12}$ &~~ -1.0842158859E+03~~
\\
$p_5$ &~~-7.9619122471E+01~~
 &~~ $p_{13}$ &~~ 4.2916504209E+02~~
\\

\hline \hline
\end{tabular}
\end{table}

\begin{figure}
\begin{center}
\includegraphics[height=3in]{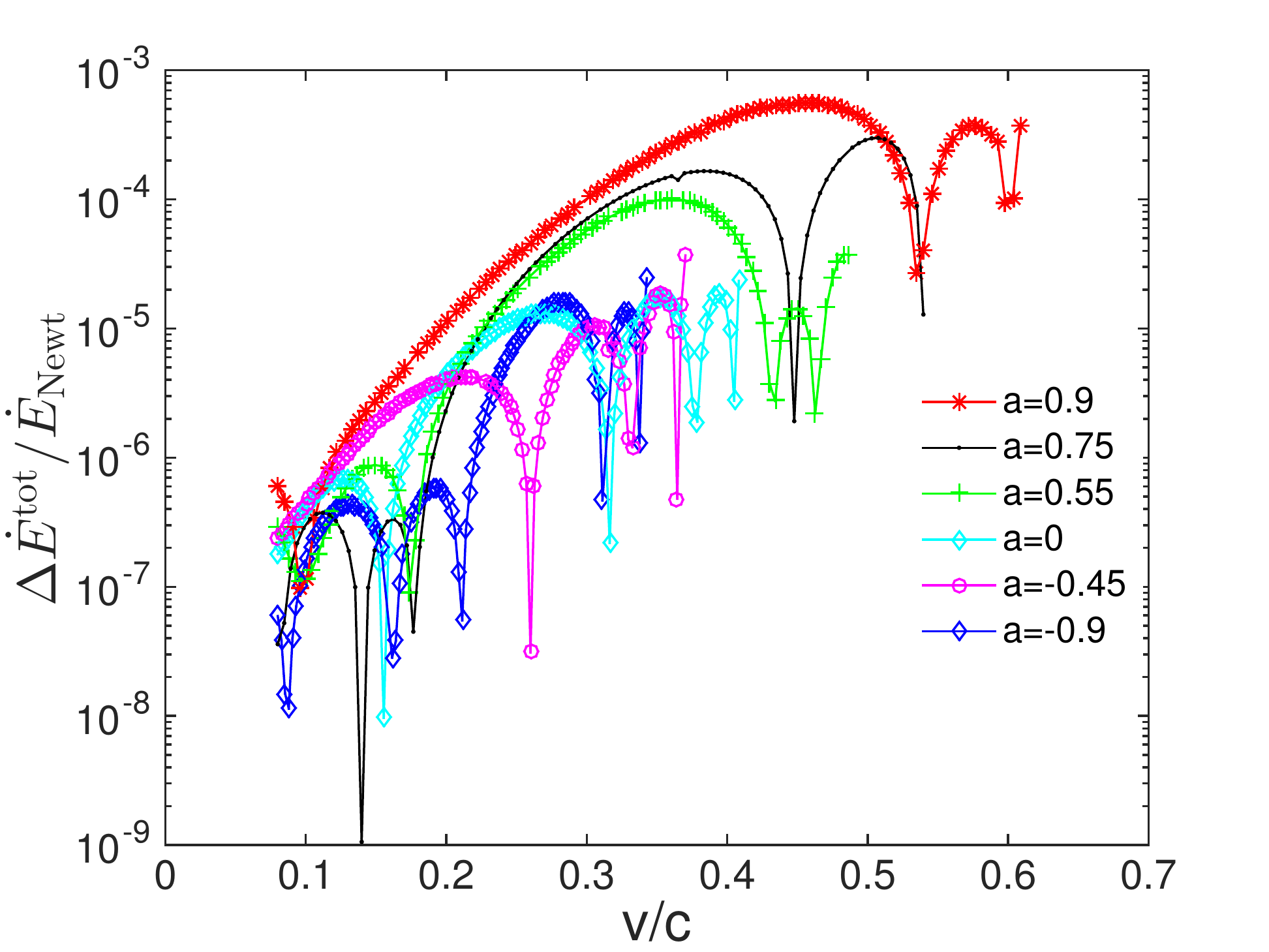}
\caption{The relative errors of the total energy fluxes which include the black hole absorption terms. $\dot{E}_{\rm Newt} \equiv 32/5 \nu^2 v^{10}$ is the leading-order (Newtonian) piece of the flux.} \label{ge8} \label{dephase}
\end{center}
\end{figure}

Theoretically, the higher the precision of the fluxes is, the more accurate evolutionary waveforms it will induce. To confirm the accuracy of the EOB model with our recalibrated waveforms, we choose a specific EMRI like the one in Ref. \cite{Yunes11}. For simplicity, we just choose such an EMRI with masses $M = 10^6 M_\odot$ and $m = 10M_\odot$ for a mass ratio of $10^5$ and it inspirals for $\sim 10^5$ rads of orbital phase depending on the spin. We evolve such EMRI with spin background of $q=0.9, ~0$ and -0.9, and start the evolution at an initial orbit radius $r_0 = 11 M$. The evolution lasts about 32 months for the small object reaches at the ISCO. Because the radius of the ISCO of in case the of $q=-0.9$ is about $9 M$(comparatively, the one in the case of $q=0.9$ is about $2.3M$), the EMRI just evolves about 8 months.

\begin{figure}
\begin{center}
\includegraphics[height=3in]{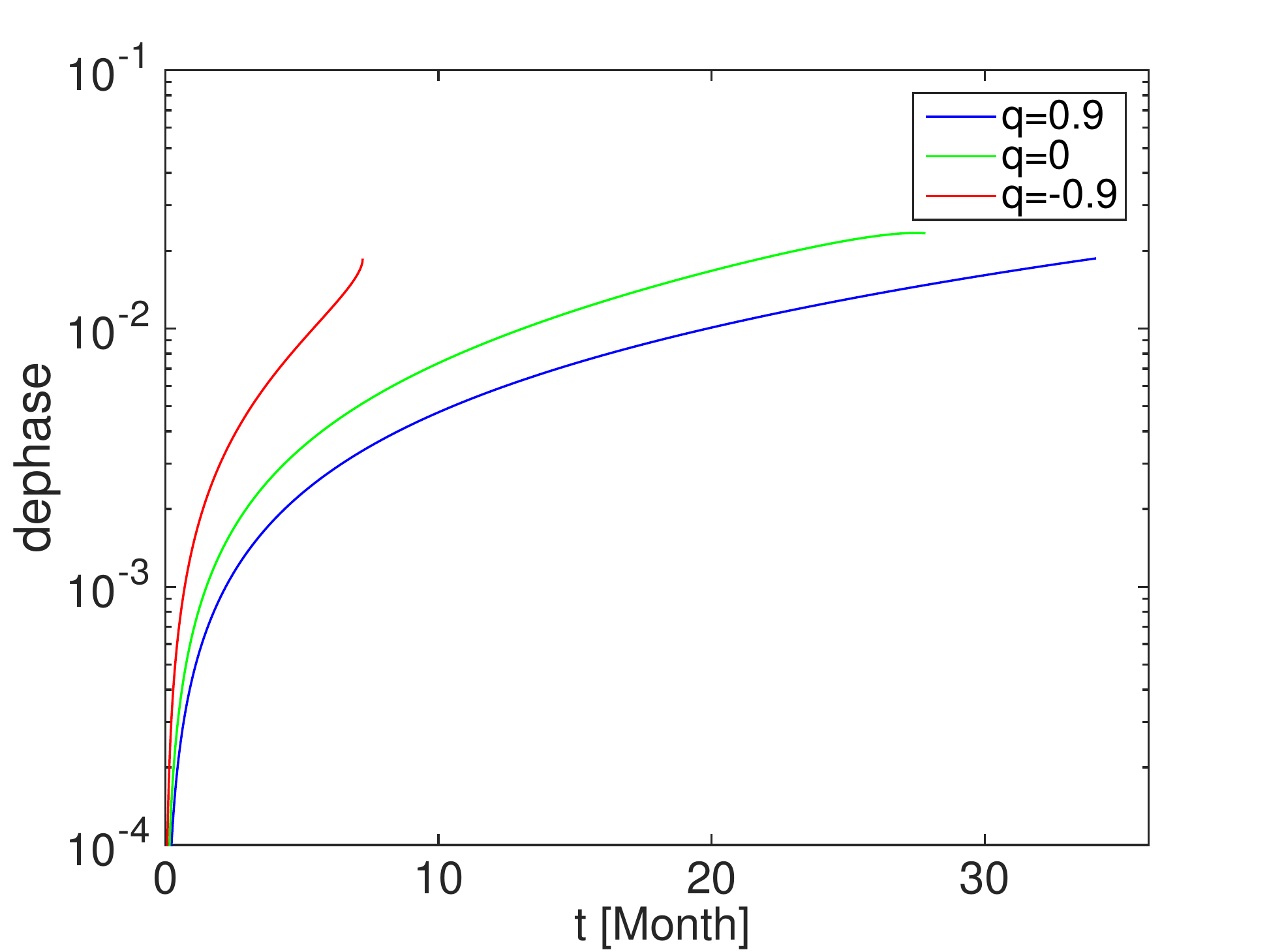}
\caption{Absolute value of the dephasing (in radians) computed in the recalibrated EOB model and the Teukolsky-based waveforms for the dominant (l, m) = (2, 2) mode. Different curves correspond to different background spin values. The $q=0$ and -0.9 cases stop earlier because they arrive at the ISCO earlier.} \label{totale}
\end{center}
\end{figure}

The dephase of dominated (2, 2) mode-waveforms calculated in the EOB frame with the newly recalibrated formalisms and the Teukolsky-based ones is shown in Fig. \ref{totale}. During the whole evolution of all spin parameters, the dephase is less than 0.1 rad. This performance should satisfy the requirement of the data analysis of the EMRIs. The reason for our better performance on the dephase lies in the high accuracy of the total energy fluxes (including black hole absorption term). We have two more orders improvement on the accuracy of the fluxes compared with the results in \cite{Yunes11}, and this induces also two orders improvement in the dephase of the 22 mode accordingly.

\section{Conclusions}
The successes of EOB model in the precise simulation of the waveforms from comparable mass-ratio binaries motivate us to develop waveform models for extreme-mass-ratio inspirals in the EOB framework. In the present work, for improving the accuracy of the factorized-resummation PN waveforms, we recalibrate all the spin-independent and -dependent coefficients of the PN waveform formalisms.

Firstly, we calibrate the global coefficients which are valid for all spin parameter $q$ by using 17 groups of numerical Teukolsky-based fluxes data with $q = $-0.9, -0.7, -0.5, -0.3, -0.1, 0.0, 0.1, 0.2, 0.3, 0.4, 0.5, 0.6, 0.7, 0.75, 0.8, 0.85, 0.9. This improves the accuracy of fluxes about one order compared with the previous results like as in \cite{Yunes11}. For getting the better waveform formalisms, we divide the spin background in four groups. For each group, we calibrate the coefficients just for this group. By this way, we improve the accuracy of the fluxes about two orders higher than the previous results.  Furthermore, we also calibrate a few of coefficients for the PN expressions of the energy-fluxes down to the horizon. This improves the accuracy of total energy fluxes greatly. All these coefficients for calculation and validation are listed in Table I-X.

With the recalibrated PN formalisms, we can accurately evolve the orbits of EMRIs in the EOB frame, and at the same time obtain the waveforms. In this way, the computation efficiency is much higher than solving the Teukolsky equation numerically. As an example, we calculate the dephasing of the 22 mode between the recalibrated waveforms and the Teukolksy-based ones for an EMRI case. As we can see, the dephasing is less than 0.1 rads even for more than 30 months' evolutionary waveforms. We think that the performance of the recalibrated waveforms in the EOB frame meet the requirement of the waveform templates of the EMRIs for space-based GW detectors. With these new waveforms, we can calculate the waveforms of EMRIs in long time-scale with accuracy and efficiency. We believe that our recalibrated formalisms can replace the Teukolsky-based waveforms in the future data analysis.

However, there are two points that may help to improve the results in the present paper again. First, one can add more $(l, ~m)$ modes. In this paper, we only consider the modes up to $l=8$. For the higher modes, one can write down the similar formalisms, then calibrate the coefficients from numerical data. Second, one may find a better way to calibrate the fluxes down to the horizon of black hole.

\section*{Acknowledgement}
 This work is supported by NSFC No. U1431120 and Key Research Program of Frontier Sciences No. QYZDB-SSW-SYS016 of CAS. WH is also supported by Youth Innovation Promotion Association CAS. We appreciate Prof. Yunes for his kind helps. 

\vfill\eject




\nocite{*}

{}
\end{document}